\documentclass[%
mathleft,%
]{an}
\usepackage{graphicx}
\usepackage{times}
\def\HI{H{\,\small I}}

\newcommand{\muJybeam}{$\mu$Jy beam$^{-1}$}
\newcommand{\msun}{{$M_\odot$}}
\newcommand{\msunyr}{{$M_\odot$ yr$^{-1}$}}
\newcommand{\kms}{$\,$km$\,$s$^{-1}$}
\begin{document}

\Pagespan{1}{}
\Yearpublication{2011}%
\Yearsubmission{2011}%
\Month{1}%
\Volume{999}%
\Issue{92}%

\title{Cold gas and the disruptive effect of a young radio jet}
\author{R. Morganti \inst{1,2}\fnmsep\thanks{Corresponding author:
  \email{morganti@astron.nl}}
\and  T. Oosterloo\inst{1,2}
\and F.M. Maccagni\inst{1,2}
\and K. Ger{\'e}b\inst{3}
\and J. B. R. Oonk\inst{1,4}
\and C.N. Tadhunter\inst{5}
}
\titlerunning{The disruptive
effect of a young radio jet}
\authorrunning{R. Morganti et al.}
\institute{
ASTRON, the Netherlands Institute for Radio Astronomy, Postbus 2, 7990 AA, Dwingeloo, The 
Netherlands.
\and
Kapteyn Astronomical Institute, University of Groningen, P.O. Box 800,
9700 AV Groningen, The Netherlands
\and 
Centre for Astrophysics \& Supercomputing, Swinburne University of Technology, Hawthorn, VIC 3122, Australia
\and
Leiden Observatory, Leiden University, P.O. Box 9513, 2300 RA Leiden, The Netherlands
\and
Department of Physics and Astronomy, University of Sheffield, Sheffield, S7 3RH, United Kingdom
}
\received{XXXX}
\accepted{XXXX}
\publonline{XXXX}
\keywords{galaxies: active - ISM: jets and outflows, radio lines: galaxies}

\abstract{
Newly born and young radio sources are in a delicate phase of their life. Their jets are fighting their way through the surrounding gaseous medium,  strongly experiencing this interaction while, at the same time, impacting and affecting the interstellar medium (ISM).  Quantifying this interplay has far reaching implications: the rate of occurrence and the magnitude of the interaction between radio jets and  ISM can have consequences for the evolution of the host galaxy. 
Despite the hostile conditions, cold gas - neutral atomic hydrogen and molecular - has been often found in these objects and can be also associated to fast outflows.   Here we present the results from two studies of \HI\ and molecular gas illustrating what can be learned from these phases of the gas. 
We first describe a statistical study of the occurrence and kinematics of  \HI\ observed in absorption with the Westerbork Synthesis Radio telescope. This allows a comparison between the properties of the gas in extended and  in compact/young radio sources. The study shows that the young radio sources not only have an higher detection rate of \HI,  but also systematically broader and  more asymmetric \HI\ profiles, most of them blueshifted. This supports the idea that we are looking at young radio jets making their way through the surrounding ISM,  which  also appears to be, on average, richer in gas than in evolved radio sources. Signatures of the impact of the jet are seen in the kinematics of the gas, but the resulting outflows may be characteristic of only the initial phase of the radio source evolution. However, even among the young sources, we identify  a population that remains undetected in \HI\ even after stacking their profiles. Orientation effects can only partly explain the result. These objects either are genuinely gas-poor or have different conditions of the medium, e.g.\ higher spin temperature. The upcoming blind \HI\ surveys which are about to start with large-field-of-view radio facilities (i.e.\ Apertif at the WSRT and ASKAP) will allow us to expand the statistics and reach even higher sensitivity with stacking techniques.
We further present the case of the radio source IC~5063 where we have used the molecular gas observed with ALMA to trace  in detail the jet impacting the ISM. The kinematics of the cold, molecular gas co-spatial with the radio plasma shows this process in action. The ALMA data reveal  a
  fast outflow of molecular gas extending along the entire radio jet ($\sim$1 kpc), with the highest outflow velocities at
  the location of the brighter hot-spot. The results can be described by a scenario of  a radio plasma jet expanding into a clumpy medium, interacting directly with the  clouds and inflating a cocoon that drives a lateral outflow into the ISM. This is consistent with the scenario proposed by numerical simulations for the expansion of a young radio jet, confirming the disruptive effect  the radio plasma jet can have.  Following this case, more  ALMA observations of nearby young radio sources will be able to confirm if this process is common, as expected, in the initial phase of the evolution of the radio source.
}
\maketitle

\section{Introduction}
A variety of interesting studies have focused on the properties of the interstellar medium (ISM) that surrounds young radio sources (i.e.\ Compact Steep Spectrum and Gigahertz Peaked Spectrum; CSS/GPS). Observing the  occurrence, distribution and kinematics of the gas gives the opportunity to investigate  the interplay between the young radio jets and the  surrounding medium. This is crucial for understanding and quantifying the impact of the radio plasma on the gas (e.g.\ by affecting its conditions, by driving ionising shocks, or by producing outflows) and of the medium on the  young radio source (e.g.\ by slowing down or even  frustrating its evolution).

\begin{figure*}[!t]
\begin{center}
\includegraphics[width=.4\textwidth]{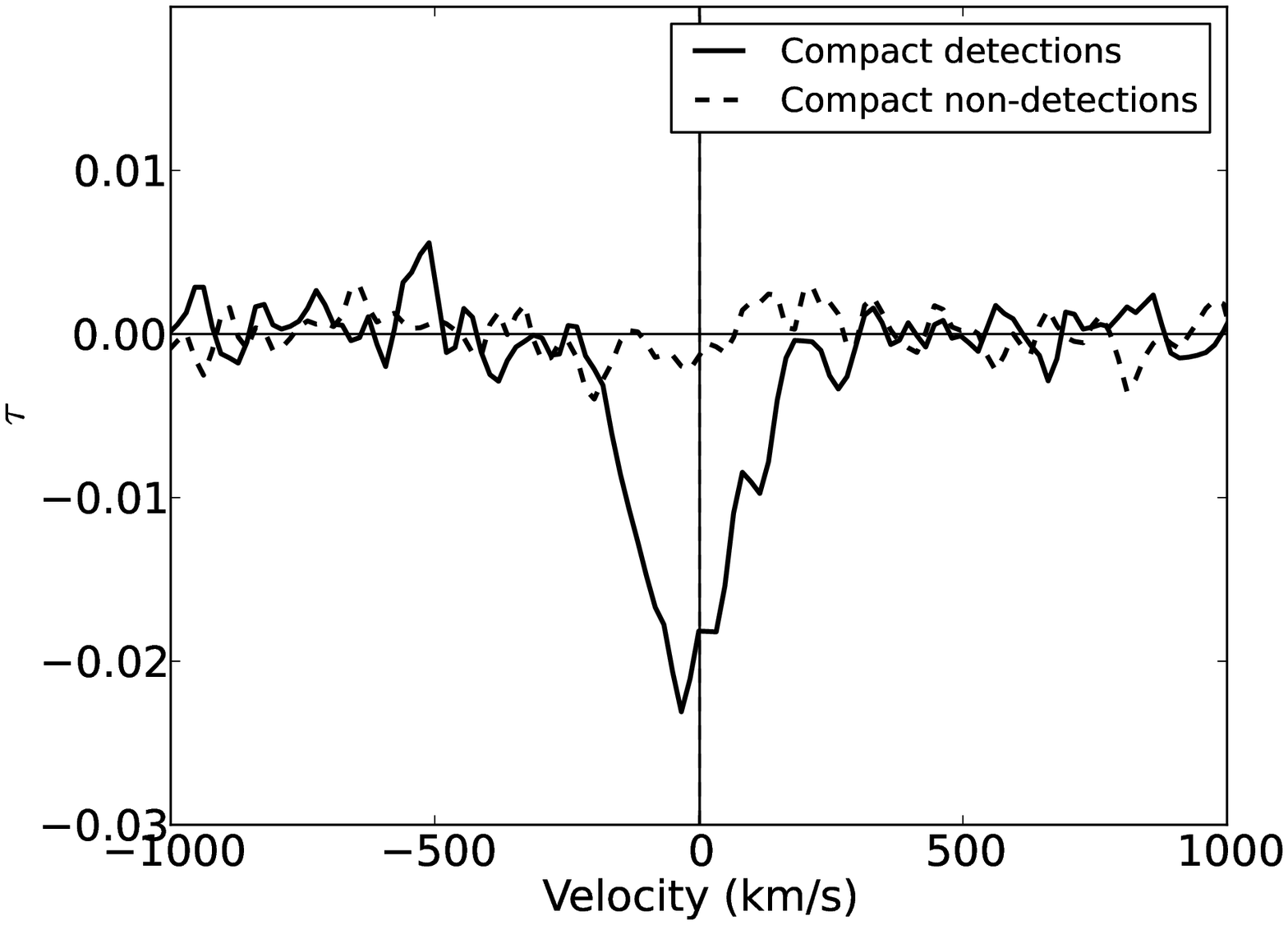}
\includegraphics[width=.4\textwidth]{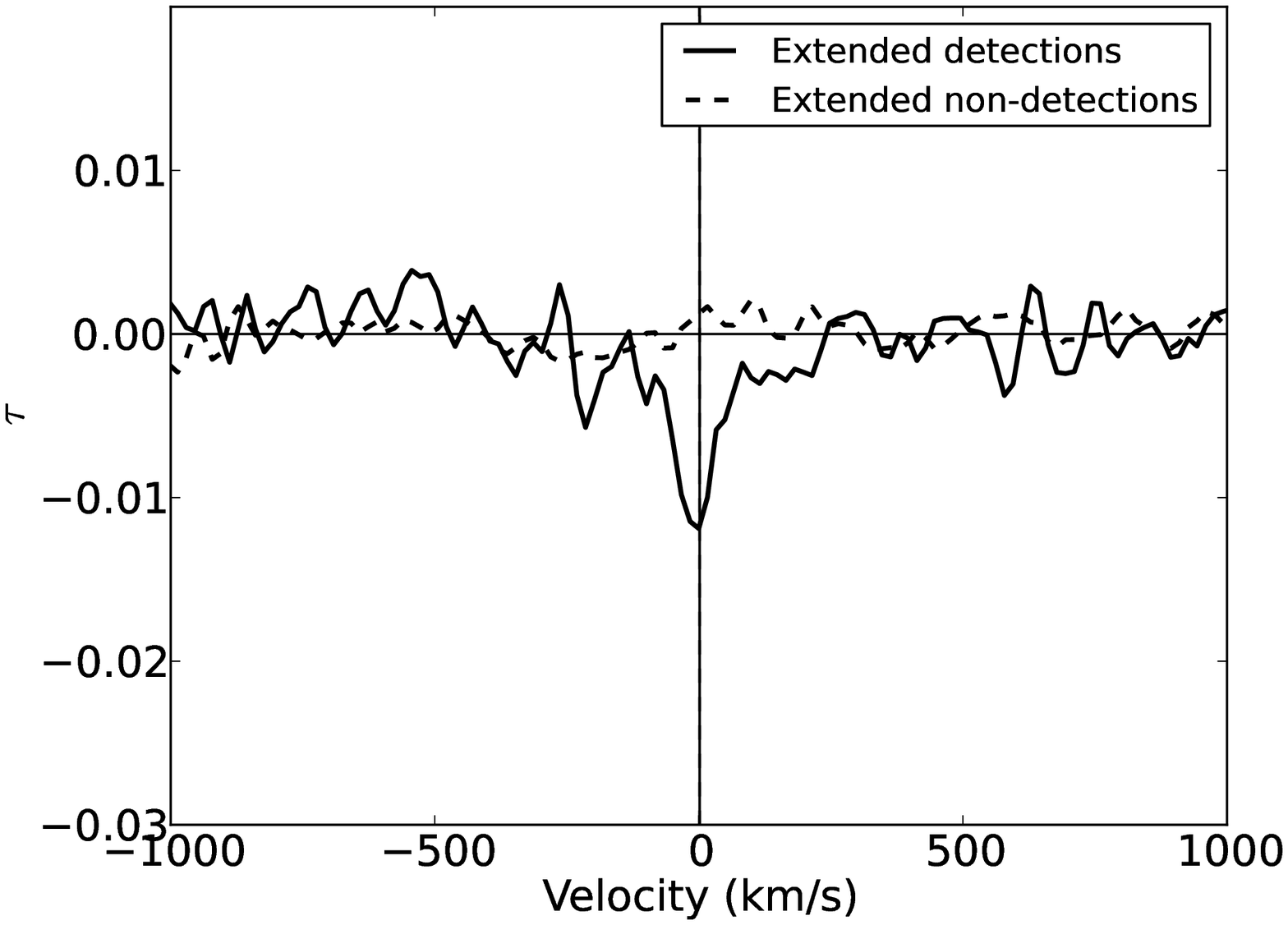}
\caption{The stacked profiles from Ger{\'e}b et al.\ (2014) of compact and  of extended sources on the left and right side respectively. Detections are indicated by dashed and dotted lines, and non-detections by solid lines.}\label{stack:CompactDetected}
\end{center} 
\end{figure*}


The study of the ionised gas and extended emission line regions (Shih et al.\ 2013;  Holt et al.\ 2008, 2009; O'Dea et al.\ 2002; Axon et al.\ 2000) has  given valuable information  by showing that the distribution, as well as the kinematics of the gas, is affected by the expanding radio source and fast outflows are often observed, albeit characterised by a low mass outflow rate.  However, despite this relatively unfriendly environment, not only ionised  but also atomic (Pihlstr{\"o}m et al.\ 2003, Vermeulen et al.\ 2003,  Chandola et al.\ 2011, Ger{\'e}b et al.\ 2015) and molecular gas (e.g.\ Mack et al.\ 2009, Dasyra \& Combes 2012, Garc\'ia-Burillo et al.\ 2007) is detected.
Recent studies have shown that the cold gas plays an important role in young radio sources, both  in connection to the fuelling (e.g.\ Maccagni et al.\ 2014) as well  as  to their evolution being  associated to jet-driven outflows (e.g.\ Morganti et al.\ 2013 and refs therein).  Given these results, it is particularly exciting that the study of cold gas is on the verge of a major expansion by taking advantage of the newly available facilities  (e.g.\ ALMA) or those that will soon start (e.g.\ SKA pathfinders and precursors), opening up new possibilities for exploring further this field. 

Some preparatory work is necessary in order to fully explore these possibilities.  Here we present results obtained by two different projects, both providing more insights on the interplay between radio plasma and ISM while illustrating the great potential of the upcoming expansion of these studies in the near future. 
The first project is preparatory for the upcoming large, blind \HI\ surveys and addresses  the occurrence of \HI\ and \HI\ outflows in young radio sources using a snap-shot survey with the Westerbork Synthesis Radio Telescope (WSRT). 
This project was motivated by the need to expand on what done so far on known samples of radio galaxies.   
The second part shows an example of a jet-ISM interaction traced by molecular gas observed in CO(2-1), illustrating the great potential of  ALMA for the study  the interaction and physical conditions of the gas connected to this process. The comparison with  theoretical models is now possible with the wealth of details provided by these data.

\section{A "zoo" of  \HI\ absorption}

The initial goal of this project was to collect a large number of short \HI\ (absorption) observations targeting radio sources to develop the infrastructure and the  automated procedures which will be necessary to handle the data of the upcoming large surveys (e.g.\ automatic detection of absorption features, characterisation of the profiles, classification in compact/extended radio sources and stacking of the profiles). To achieve this, we observed with the WSRT sources selected from cross-correlating FIRST radio sources with $S_{\rm 1.4 GHz}>$ 30 mJy  and SDSS galaxies with an available spectroscopic redshift  $z<0.2$ (i.e.\ \HI\ falling into the WSRT  observing  band).  The sources were observed for only a few hours (i.e.\ no full synthesis) and we have by now more than  200 sources observed. Here we focus on the discussion of the first 120 ($S_{\rm 1.4 GHz}>$ 50 mJy), as presented in Ger{\'e}b et al.\ 2014, 2015. Our study obtained the largest sample and covers lower radio fluxes than any study done so far for  \HI\ absorption. 
The range in radio power of the sources is $10^{23}$ - $10^{26}$ W/Hz.
In this study we developed an automatic procedure to separate "extended" from "compact" sources (see Ger{\'e}b et al.\ 2014 for details). The effectiveness of the methods was confirmed by checking known sources, like the  COmpact RAdio sources at Low redshift sample  (CORALZ, Snellen et al.\ 2004, de Vries et al.\ 2010) which  indeed appear among our "compact" sources.

One of the most surprising results has been the high detection rate of  \HI\ absorption, about 30\%, despite the observations being only snap-shot.  For the detected sources, we used an automatic routine (BusyFunction, see  Westmeier et al.\ 2014) to characterize the shapes and asymmetries of the \HI\ profiles and  to quantify the kinematical properties (see Ger{\'e}b et al.\ 2015 for details). 
The observed profiles show a broad variety of properties in terms of optical depth, width and shape.  Based on the width they  can be separated in three groups, likely representing physically different \HI\ structures. The first group consists of narrow, single components ($FWHM <100$ \kms), the second group of broader profiles with two (or more) blended components ($100< FWHM <200$\kms). Among the third group are the broadest profiles ($FWHM >200$ \kms)  often with distinct components. While for the first two the \HI\ absorption may originate from small isolated clouds (narrow lines) or rotating disks at different scales,  for the last group the kinematics of the gas is too extreme and suggests the presence of unsettled gas (see below).


\begin{figure}
\centering
\includegraphics[width=.40\textwidth]{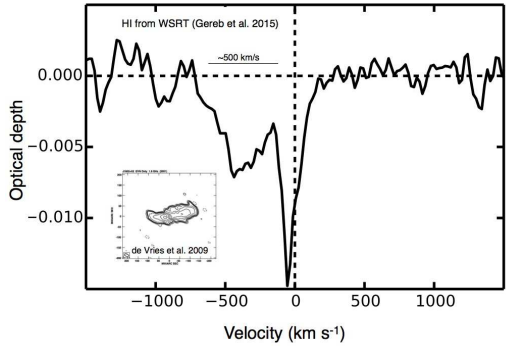}
\caption{Example of a newly detected \HI\ outflow in one young radio source. The object, 4C~52.37 is part of the CORALZ sample (de Vries et al.\ 2009). The dashed line indicate the systemic velocity and the broad part of the \HI\ profile is clearly blueshifted indicating a fast outflow.}
\label{fig:outflow}
\end{figure}


\section{\HI\ tracing young radio sources clearing their way}

In the context of the theme of this workshop, one of the most relevant results is that  compact (young) sources have  a higher detection rate of \HI\ absorption ($\sim 55$\% for the CORALZ group) compared to extended sources, confirming - and expanding  to  lower radio power - what  was suggested by previous studies  (e.g.\ Vermeulen et al.\ 2003, Chandola et al.\ 2011). 
In addition to this, we find clear differences between the characteristics of the  \HI\ profiles in extended and compact radio galaxies. The difference can already be seen in the average profiles obtained by the stacking (see Fig.\ 1). 
The {\sl detected} compact sources (Fig.\ 1 left) have higher average optical depth ($\tau \sim 0.025$ compared to $\tau \sim  0.01$ for extended sources). Furthermore, compact radio galaxies have \HI\ profiles that are broader, more asymmetric, more blueshifted than extended radio galaxies and, in particular,  most of them belong to the third class mentioned above (Ger{\'e}b et al.\ 2015).
As further support to this, we find  \HI\ outflows (identified by broad, blueshifted profiles) to be more  common in young sources. Interestingly, new cases of \HI\  outflows were found (see Fig.\ 2) despite the shallow observations. We derive that  $\sim 15$\% of \HI\ detections show outflows, i.e.\ $\sim 5$\% of all the radio sources of the sample. Thus, if a phase of outflow is characteristic of the early stage of {\sl every} radio source, then it should last not more than a few Myr. 

However, Fig.\ 1 also shows a dichotomy that appears to be present in both groups of sources.  The stacking of \HI\ undetected sources results in an upper limit to the \HI\ for both extended and compact groups of sources  despite the significant improvement in sensitivity.
This is an intriguing result  which  may suggest the presence of orientation effects consistent with the fact that we expect a number of detections due to circumnuclear (or larger scale) disks and only for objects where the disk is relatively edge-on we can see absorption. This is partly supported by the fact that the most highly inclined objects (at $b/a < 0.6$, identified  using the minor-to-major axis ratio in the $r$ band from  SDSS)  all show \HI\ absorption.
However, this does not appear to be the only explanation and, in particular, it does not explain why the dichotomy is present also in young radio sources with pc-scale sizes. According to the  correlation between column density and linear size suggested by Pihlstr\"om et al.\ (2003), small radio sources would  be expected to be embedded in a medium with high column density. We do not confirm this correlation and, as shown in Fig.\ 3 (that includes only CSS/GPS sources from our sample), a number of small radio galaxies, including High Frequency Peakers (Orienti et al.\ 2006),  do show only upper limits to the \HI\ column density. Thus, other effects may be the cause of the observed dichotomy. 
We may be  seeing  a population of objects that has a genuinely low column density medium, or where the gas is dominated by a warmer, high $T_{\rm spin}$ component (Kanekar et al.\ 2011), therefore more difficult to be detected at the typical optical depth allowed by our observations.
Indeed, Fig.\ 3 shows that in some cases (like PKS B1718-639; Maccagni et al.\ 2014) sensitive observations are needed in order to reveal \HI\ in absorption.
Furthermore, investigating the possibility of differences in the physical conditions of the gas (including covering factor, as discussed by Curran et al.\  2013) is part of the work presented by Ostorero et al.\  (these proceedings) that also illustrates the need for high resolution (VLBI) observations to trace the distribution of the \HI\ absorption. 
 

\begin{figure}
\centering
\includegraphics[width=.30\textwidth, angle=-90]{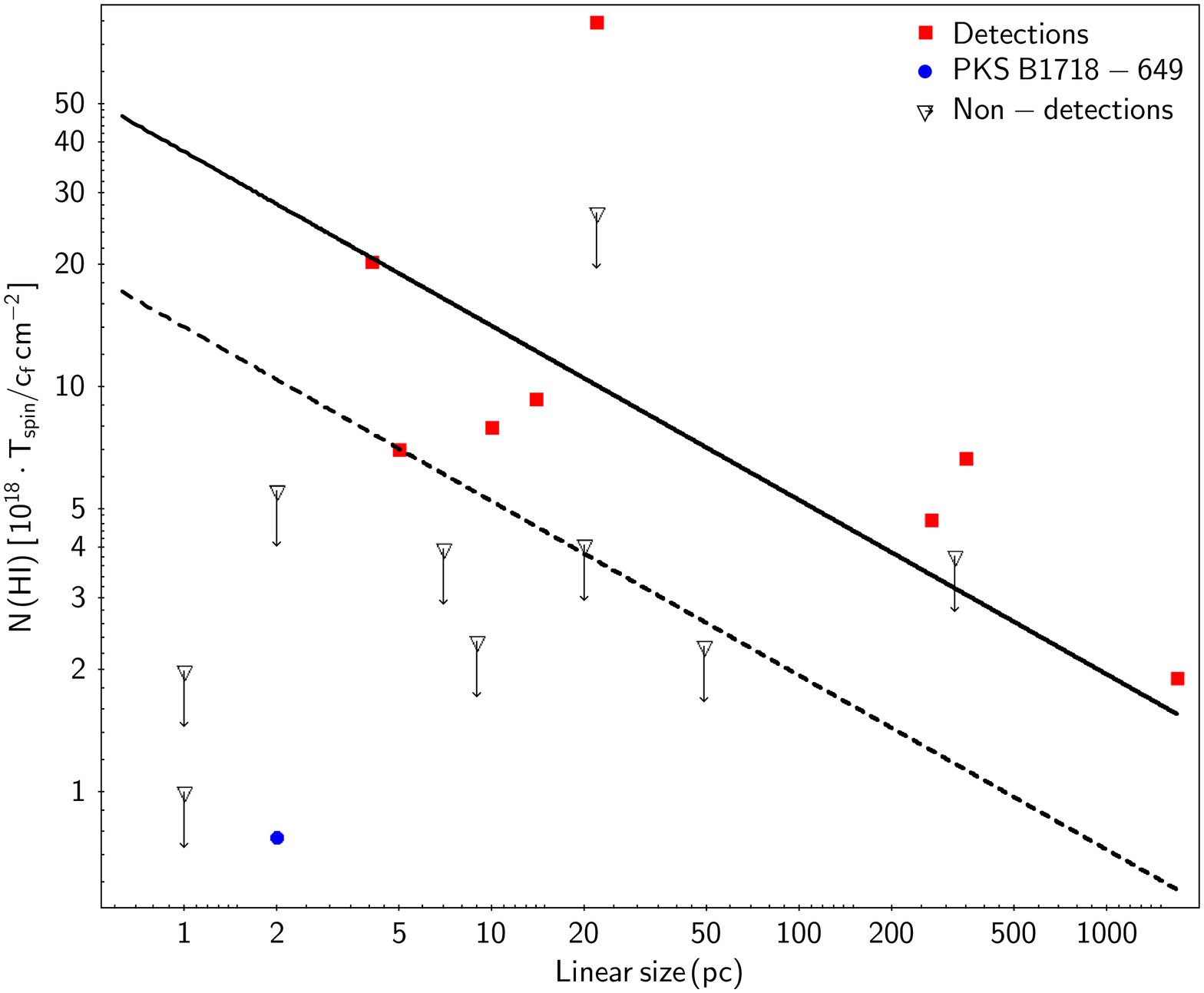}
\caption{Radio source size vs. column density in GPS and CSS sources from the CORALZ sample (red circles) from Ger{\'e}b et al.\ 2015, HFP from Orienti et al.\ (2006) and PKS B1718--649 from Maccagni et al.\ (2014). The lines represent the relations (solid line: only for detections; dashed line: detections and upper limits)  proposed for their sample by Pihlstr\"om et al.\ (2003).}
\label{fig:outflow}
\end{figure}


\section{The effect of a young radio jet  traced by the molecular gas: IC~5063}

The \HI\ results support the idea that young radio sources strongly disturb the ISM in which they are embedded. Understanding how the interaction between their jets and the surrounding medium proceeds is crucial in order to quantify the consequences on the evolution of the host galaxy. 
Predictions about this process have been made using numerical simulations (see e.g.\ Wagner \& Bicknell 2011, 2012).  So far, only in a small group of young (CSS/GPS) radio sources the molecular gas has been studied in detail and only a handful shows disturbed  kinematics of the gas at the sensitivity of the available observations (see Dasyra \& Combes 2012, Garc\'ia-Burillo et al.\ 2007). However, this number is going to greatly expand using ALMA. Here we show an example of what ALMA can provide.
Although IC~5063 may not be considered a "standard" CSS\footnote{IC~5063 has a size of about 1~kpc  and  steep spectrum radio emission  down to 230~GHz but no clear low frequency turnover although flattening at low frequency (150~MHz measurement from MWA, Callingham Joseph private comm.)  broadly consistent with what expected, for  the size of the source, from synchrotron self-absorption  (Fanti et al.\ 1990).},  it represents a  young radio source expanding in a rich medium and, therefore, is a perfect test case to investigate the aftermath of this process. 

\begin{figure}
\begin{center}
\includegraphics[width=.5\textwidth]{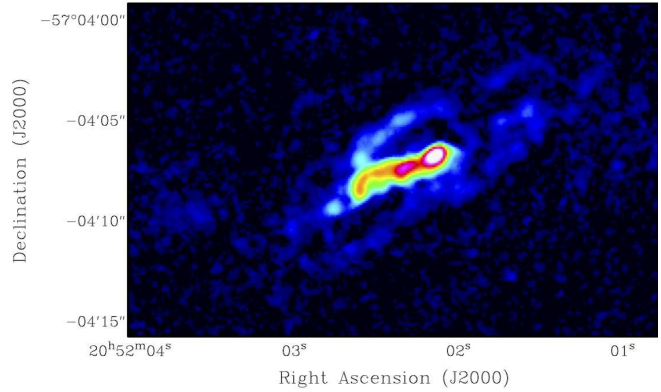}
\caption{Total intensity image representing the distribution of the CO(2-1) in
  IC 5063 and showing the striking brightness contrast between the inner,
  bright CO and the fainter outer disk. }
\end{center} 
\label{fig:TotCO}
\end{figure}

\begin{figure}
\begin{center}
\includegraphics[width=.5\textwidth]{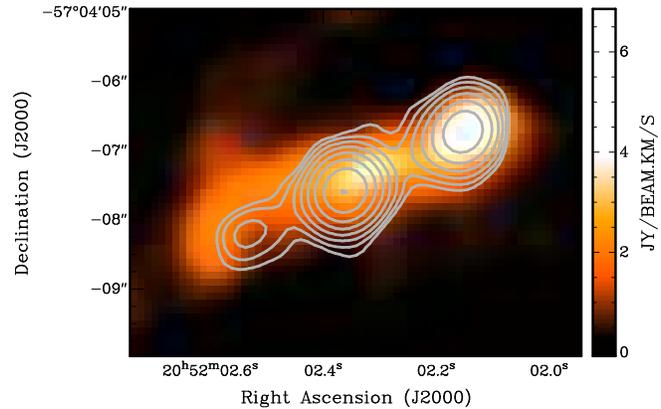}
\caption{Contours of the 230~GHz  continuum emission superposed to the central
  region of the total intensity of the CO(2-1) illustrating the spatial
  correlation between the two, from Morganti et al. (2015). Contour levels are 90, 180, 360, 720, 1440, 2880, 5760 and 15120 \muJybeam.}
\end{center} 
\label{fig:ContCO}
\end{figure}

IC~5063 was observed in CO(2-1) during Cycle 1 with ALMA. A complete description of the results can be found in Morganti et al.\ (2015).
In Fig.\ \ref{fig:TotCO} the total intensity of the CO(2-1) is shown.  The  zoom-in  of the central region  overlaying  the mm continuum image is shown in Fig.\ \ref{fig:ContCO}. The spatial resolution obtained is $\sim 0.5$ arcsec and the r.m.s.\ noise per channel in the CO cube is  0.3 mJy beam$^{-1}$ for a resolution of 20 km s$^{-1}$.
While the large-scale structure follows a regularly rotating  disk, {\sl the gas in the inner regions is strongly affected by the radio jet}. This can be clearly seen in Fig.\ \ref{fig:outflow} and from the continuum emission nicely   being co-spatial with the brighter (and  kinematically most disturbed)  inner-region detected in CO (see also Fig.\ \ref{fig:ContCO}). Thus, the ALMA data illustrate in an unprecedented way how the jet-ISM interaction  develops and affects the medium. 
Furthermore, they confirm the multi-phase nature of the gaseous outflow in IC~5063, adding the cold molecular gas to the \HI, warm molecular and ionised gas  (see Tadhunter et al.\ 2014 and refs therein).  

The observed distribution and kinematics of the cold molecular gas suggests that  the radio plasma jet is driving the bulk of the gas outflow.   Therefore, we have developed a simple model to describe the kinematics of the gas, following the scenario presented in the numerical simulations of Wagner et al.\ (2011, 2012). These simulations describe the effects of a newly formed radio jet when moving though a {\sl dense clumpy medium}. This medium forces the jet to find the path of least resistance, while interacting and gradually dispersing the dense clouds away from the jet axis.  In this way, clouds can be accelerated to high
velocities and over a wide range of directions. Along the path of the jet, a turbulent cocoon of expanding gas forms, moving away from the jet axis. 
In our toy-model, we introduce  lateral expansion of the gas, away from the jet axis, and the interaction with {\sl rotating} gas. The full description of the results and of the model can be found in Morganti et al.\ (2015).  Despite its simplicity, the model provides a good first-order description of the observations.
The origin of the cold gas is likely related  to the efficient cooling of the gas after the shock produced by the jet: cold molecular gas would represent the final product of this cooling process, while warm molecular and \HI\ would be  intermediate (and less massive) phases.  


We find that the mass of the outflowing molecular gas is between $1.9$ and $4.8 \times 10^7$ \msun, of which between $0.5$ and $1.3 \times 10^7$ \msun\ is associated  with the fast outflow at the location of the W lobe. These amounts are much larger that what found in the other phases of the gas.  Using these values, we derive a mass outflow rate  ($\dot{M} = M/\tau_{\rm dyn}$) in the range 12 to 30 \msunyr. 
Interestingly, the outflow does not appear to be fast enough to remove significant amounts of gas from the galaxy. Most likely, the main effect is to 
relocate the gas in a process comparable to a Ògalactic fountainÓ.  Furthermore, the injection of large amounts of energy in the gas disk may have a similar effect as increasing the turbulence of the gas, a process that is also thought to inhibit star formation (see the case of 3C~326, Guillard et al. 2015).


\section{Cold gas tracing the impact of radio jets: a bright future ahead}
 
The detection rate of \HI\  absorption obtained from our snap-shot observations is very  promising for upcoming surveys. The detection rate does not appear to depend on radio flux density, suggesting that future surveys may be able to trace  \HI\ (in absorption) against quite faint radio sources.  Thus, if this is confirmed, the surveys planned by upcoming radio telescopes with large Field-of-View (like Apertif on the WSRT, Oosterloo et al.\ 2009 and ASKAP Johnston et al.\ 2008) will provide a large number of new detections to investigate the properties of the \HI\ in radio sources as function of a variety of properties (e.g.\  radio power, size, age, environment etc.). A taste of the possibilities and potential of these new instruments has been shown in Allison et al.\ (2015) reporting new ASKAP results.
These surveys will be followed by even more ambitious plans using the SKA (Morganti, Sadler \& Curran 2015).

Our results have also shown that VLBI follow-up observations are extremely important  in order to learn about the distribution of the absorber and its characteristics. They have been  proven to be crucial for identifying the location of outflows (Morganti et al.\ 2013).  Follow-up observations  of some of the new detections described in this paper are already in progress. The need to know more about the characteristics of the covering factor and $T_{\rm spin}$ of the absorber requires to trace its distribution with high spatial resolution  (see also Ostorero et al.\ these Proceedings).
 
The  potential of ALMA has been clearly demonstrated by the results on IC~5063. These observations seems to confirm the predictions from the numerical simulations  (Wagner  et al.\ 2011, 2012). According to these models, the radio jet has a higher impact when expanding  in a clumpy medium, affecting a large volume of the bulge around it. This scenario will need to be confirmed with detailed studies of more objects. Finally, the idea that it is in the initial phases (or restarted phase) of their life that radio jets have the largest impact on the ISM seems to be supported by a number of results.This will need to be further verified by tracing the distribution and kinematics of the cold gas in objects in different phases of their evolution.  Also in this, we expect a major role of ALMA taking advantage of the sensitivity and long baselines to spatially resolve the distribution of the gas and trace its kinematics. 

In summary, a lot of exciting possibilities are becoming available between now and the next CSS/GPS workshop!


\begin{figure}
\centering
\includegraphics[width=\hsize, keepaspectratio]{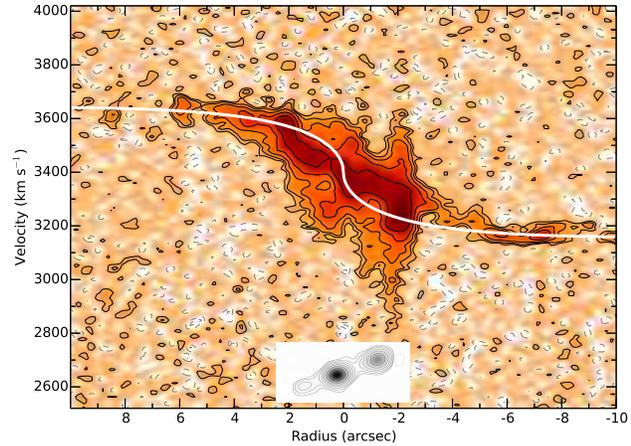}
\caption{Integrated (over 2$^{\prime\prime}$ perpendicular to the major axis) position-velocity map taken along the major axis of IC~5063, from Morganti et al.\ 2015. The white line gives the rotation curve we derived from the photometry of Kulkarni et al.\ 1998 and  illustrates the expected kinematics of gas following regular rotation.}
\label{fig:PVCO}
\end{figure}


\acknowledgements
The research leading to these results has received funding from the European Research Council under the European Union's Seventh Framework Programme (FP/2007-2013) / ERC Advanced Grant RADIOLIFE-320745.
RM would like to thank the organisers and RadioNet for the support to this inspiring workshop.

%

\begin{thebibliography}{}

\bibitem[\protect\citeauthoryear{Allison et  al.}{2015}]{2015arXiv150301265A} Allison J.R., et al.: 2015, MNRAS subm.,
arXiv:1503.01265 

\bibitem[]{}Axon, D. J., Capetti, A., Fanti, R., et al. 2000, AJ 120, 2284

\bibitem[Chandola et al.(2011)]{Chandola} Chandola, Y., Sirothia, S., \& Saikia, D.: 2011, MNRAS 418, 1787 

\bibitem[]{}Curran, S. J., Allison, J. R., Glowacki, M., Whiting, M. T.,\& Sadler, E. M. 2013,
MNRAS 431, 3408


\bibitem[\protect\citeauthoryear{Dasyra 
\& Combes}{2012}]{2012A&A...541L...7D} Dasyra K.~M., Combes F., 2012, A\&A, 541, L7 

\bibitem[]{}Fanti, R., et al.: 1990, A\&A 231, 333

\bibitem[]{}Garc\'ia-Burillo, S. et al.	2007A\&A 468, L71

\bibitem[Gereb et al.(2014a)]{Gereb2014a} Ger{\'e}b K., Morganti R., Oosterloo T.~A.: 2014, A\&A 569, AA35

\bibitem[Gereb et al.(2014b)]{Gereb2014b} Ger{\'e}b K., Maccagni F., Morganti  R., Oosterloo T.:  2015, A\&A 575, 44

\bibitem[]{}Guillard P., et al. 2015: A\&A, 547, 32

\bibitem[]{} Gupta, N., Salter, C. J., Saikia, D. J., Ghosh, T., \& Jeyakumar, S. 2006, MNRAS,
373, 972

\bibitem[]{}Holt, J., Tadhunter, C. N., \& Morganti, R. 2008, MNRAS 387, 639

\bibitem[]{}Holt, J., Tadhunter, C. N., \& Morganti, R. 2009, MNRAS 400, 589

\bibitem[\protect\citeauthoryear{Johnston et 
al.}{2008}]{2008ExA....22..151J} Johnston S., et al., 2008, ExA, 22, 151 

\bibitem[\protect\citeauthoryear{Kanekar, Braun, 
\& Roy}{2011}]{2011ApJ...737L..33K} Kanekar N., Braun R., Roy N., 2011, ApJ, 737, L33 

\bibitem[Kulkarni et al.(1998)]{Kulkarni1998} Kulkarni, V.P. et al.: 1998, ApJ 492, L121 

\bibitem[\protect\citeauthoryear{Maccagni et 
al.}{2014}]{2014A&A...571A..67M} Maccagni F.~M., Morganti R., Oosterloo T.~A., Mahony E.~K., 2014, A\&A, 571, A67 

\bibitem[\protect\citeauthoryear{Mack et al.}{2009}]{2009AN....330..217M} 
Mack K.-H., Snellen I.~A.~G., Schilizzi R.~T., de Vries N., 2009, AN, 330, 
217
 

\bibitem[\protect\citeauthoryear{Morganti et 
al.}{2015}]{2015arXiv150507190M} Morganti R., Oosterloo T., Oonk J.~B.~R., 
Frieswijk W., Tadhunter C.: 2015, A\&A 580, 1 


\bibitem[\protect\citeauthoryear{Morganti, Sadler, 
\& Curran}{2015}]{2015aska.confE.134M} Morganti R., Sadler E.~M., Curran S., 2015, aska.conf, 134 

\bibitem[\protect\citeauthoryear{Morganti et 
al.}{2013}]{2013Sci...341.1082M} Morganti R., Fogasy J., Paragi Z., 
Oosterloo T., Orienti M., 2013 Science, 341, 1082 

\bibitem[\protect\citeauthoryear{O'Dea et al.}{2002}]{2002AJ....123.2333O} 
O'Dea C.~P., et al., 2002, AJ 123, 2333 

\bibitem[\protect\citeauthoryear{Oosterloo et 
al.}{2009}]{2009wska.confE..70O} Oosterloo T., Verheijen M.~A.~W., van 
Cappellen  et al.2009, Proceedings of Wide Field Astronomy \& Technology for the SKA
{http:/\slash{}pos.sissa.it\slash{}cgi-bin/reader/conf.cgi?confid=132, id.70}

\bibitem[\protect\citeauthoryear{Orienti, Morganti, 
\& Dallacasa}{2006}]{2006A&A...457..531O} Orienti M., Morganti R., Dallacasa D., 2006, A\&A, 457, 531 

\bibitem[\protect\citeauthoryear{Pihlstr{\"o}m, Conway, 
\& Vermeulen}{2003}]{2003A&A...404..871P} Pihlstr{\"o}m Y., Conway J., Vermeulen R., 2003, A\&A, 404, 871 

\bibitem[\protect\citeauthoryear{Pihlstr{\"o}m, Conway, 
\& Vermeulen}{2003}]{2003PASA...20...62P} Pihlstr{\"o}m Y., Conway J., Vermeulen R., 2003, PASA, 20, 62 

\bibitem[\protect\citeauthoryear{Shih, Stockton, 
\& Kewley}{2013}]{2013ApJ...772..138S} Shih H.-Y., Stockton A., Kewley L., 2013, ApJ 772, 138 

\bibitem[Snellen et al.(2004)]{Snellen2004} Snellen, I.~A.~G., Mack, K.-H., Schilizzi, R.~T., \& Tschager, W.: 2004, MNRAS 348, 227 

\bibitem[Tadhunter et al.(2014)]{Tadhunter2014} Tadhunter C., Morganti R., Rose M., Oonk J.~B.~R., Oosterloo T. 2014, Nature 511, 440;

\bibitem[\protect\citeauthoryear{Vermeulen et 
al.}{2003}]{2003A&A...404..861V} Vermeulen R.~C., et al., 2003, A\&A, 404, 861 

\bibitem[de Vries et al.(2009)]{deVries} de Vries, N., Snellen, I.~A.~G., Schilizzi, R.~T., Mack, K.-H., \& Kaiser, C.~R.: 2009, A\&A 498, 641

\bibitem[Wagner \& Bicknell(2011)]{Wagner2011} Wagner A.~Y., Bicknell G.~V.: 2011, ApJ 728, 29 

\bibitem[Wagner et al.(2012)]{Wagner2012} Wagner A.~Y., Bicknell G.~V., Umemura M.: 2012, ApJ 757, 136 

\bibitem[]{}Westmeier, T., Jurek, R., Obreschkow, D., Koribalski, B. S., Staveley-Smith, L.
2014, MNRAS 438, 1176

\end{thebibliography}
%



\end{document}